\documentclass[aps,prl,twocolumn,showpacs,superscriptaddress,groupedaddress]{revtex4}  
\usepackage{graphicx}  
\usepackage{dcolumn}   

\usepackage{bm}        
\usepackage{amssymb}   
\usepackage[caption=false]{subfig}
\def\fiverm{\footnotesize}
\input prepictex.tex
\input pictex.tex
\input postpictex.tex

\usepackage[usenames,dvipsnames,svgnames]{xcolor}

\def\2;{\;\;}

\def\Ref#1{(\ref{#1})}

\def\C#1{{\mathcal #1}}

\def\Sfrac#1#2{\hbox{\large $\frac{#1}{#2}$}}

\def\LB{\left(}         \def\RB{\right)}

\def\lcl{\left\lceil}  \def\rcl{\right\rceil}
\def\lfl{\!\left\lfloor} \def\rfl{\right\rfloor\!}

\def\vv{{\;\hbox{\Large $|$}\;}}

\def\thin{ {\hspace{0.75pt}} }

\def\tiny#1{\scalebox{0.75}{#1}}

\definecolor{blue}{rgb}{0,0.18,0.39}
\definecolor{RoyalBlue}{rgb}{0,0.2,0.7}

\definecolor{Maroon}{cmyk}{0,0.87,0.68,0.62}
\definecolor{Brown}{rgb}{0.7,0.3,0}
\definecolor{Navy}{rgb}{0.3,0.0,0.4}
\definecolor{Red}{cmyk}{0,1,1,0}
\definecolor{BrickRed}{cmyk}{0.16,0.89,0.61,0.02}
\definecolor{DarkRed}{cmyk}{0,1,1,0.5}
\definecolor{DarkBlue}{cmyk}{1,1,0,0.2}
\definecolor{DarkGreen}{cmyk}{1,0,1,0.4}
\definecolor{Green}{cmyk}{1,0,1,0}
\definecolor{DarkBrown}{cmyk}{0,0.81,1,0.6}
\definecolor{OrangeRed}{cmyk}{0,1,0.87,0}
\definecolor{RedOrange}{cmyk}{0,0.77,0.87,0}
\definecolor{Orange}{cmyk}{0,0.61,0.87,0}
\definecolor{Offwhite}{rgb}{.8,0.9,.8}
\definecolor{Offwhite2}{cmyk}{.04,.02,.01,0}
\definecolor{Tan}{rgb}{0.82,0.70,0.55}
\definecolor{Blue}{rgb}{0,0,1}
\definecolor{RoyalBlue}{rgb}{0.25,0.41,0.88}
\definecolor{Sepia}{rgb}{0.37,0.14,0.07}
\definecolor{myblue}{cmyk}{0.025,0.05,0,0}
\definecolor{Mahogany}{cmyk}{0.18,0.87,1,0.08}

\definecolor{green1}{cmyk}{0.25,0,0.76,0}
\definecolor{green2}{cmyk}{0.25,0,0.76,0.07}
\definecolor{green3}{cmyk}{0.25,0,0.76,0.20}
\definecolor{green4}{cmyk}{0.25,0,0.75,0.30}
\definecolor{green5}{cmyk}{0.25,0,0.75,0.40}
\definecolor{green6}{cmyk}{0.25,0,0.75,0.50}

\definecolor{B02}{cmyk}{0,0.14,0.22,0.12}
\definecolor{B03}{cmyk}{0,0.16,0.26,0.16}
\definecolor{B04}{cmyk}{0,0.19,0.28,0.19}
\definecolor{B05}{cmyk}{0,0.25,0.32,0.25}
\definecolor{B06}{cmyk}{0,0.31,0.36,0.31}
\definecolor{B07}{cmyk}{0,0.37,0.40,0.37}
\definecolor{B08}{cmyk}{0,0.46,0.46,0.46}
\definecolor{B09}{cmyk}{0,0.55,0.52,0.54}
\definecolor{B10}{cmyk}{0,0.69,0.61,0.62}
\definecolor{B11}{cmyk}{0,0.78,0.70,0.68}
\definecolor{B12}{cmyk}{0,0.93,0.85,0.60}
\definecolor{B13}{cmyk}{0.25,1,0.6,0.50}
\definecolor{B14}{cmyk}{0.5,1,0.30,0.40}
\definecolor{B15}{cmyk}{0.75,1,0,0.30}

\definecolor{C02}{cmyk}{0,0.22,0.14,0.12}
\definecolor{C03}{cmyk}{0,0.26,0.16,0.16}
\definecolor{C04}{cmyk}{0,0.28,0.19,0.19}
\definecolor{C05}{cmyk}{0,0.32,0.25,0.25}
\definecolor{C06}{cmyk}{0,0.36,0.31,0.31}
\definecolor{C07}{cmyk}{0,0.40,0.37,0.37}
\definecolor{C08}{cmyk}{0,0.46,0.46,0.46}
\definecolor{C09}{cmyk}{0,0.52,0.55,0.54}
\definecolor{C10}{cmyk}{0,0.61,0.69,0.62}
\definecolor{C11}{cmyk}{0,0.70,0.78,0.68}
\definecolor{C12}{cmyk}{0,0.85,0.93,0.60}
\definecolor{C13}{cmyk}{0.25,0.60,1,0.50}
\definecolor{C14}{cmyk}{0.5,0.30,1,0.40}
\definecolor{C15}{cmyk}{0.75,0,1,0.30}

\begin{document}

\widetext
\leftline{Version 1.0 as of \today}

\title{Numerical estimates of square lattice star vertex exponents}
\author{S Campbell}
\affiliation{Department of Statistics, 
University of Toronto, Toronto, Ontario M3J~4S5, Canada\\}
\author{EJ Janse van Rensburg}
\affiliation{Department of Mathematics and Statistics, 
York University, Toronto, Ontario M3J~1P3, Canada\\
Email: rensburg@yorku.ca\\}

\date{\today}

\begin{abstract}
We implement parallel versions of the GARM and Wang-Landau algorithms
for stars and for acyclic uniform branched networks in the square lattice.
These are models of monodispersed branched polymers, and we estimate the 
star vertex exponents $\sigma_f$ for $f$-stars, and the entropic exponent
$\gamma_\C{G}$ for networks with comb and brush connectivity in
two dimensions.  Our results verify the predicted (but not rigorously proven) 
exact values of the vertex exponents and we test the scaling relation \cite{D86}
$$ \gamma_{\C{G}}-1 = \sum_{f\geq 1} m_f \, \sigma_f $$
for the branched networks in two dimensions.
\end{abstract}

\maketitle


\section{Introduction}

The vertex exponents of lattice star models of monodispersed branched
polymers have been studied since the 1970s.  These exponents have
been estimated numerically in numerous studies 
\cite{LWWMG85,WLWG86,WGLW86,BK89,OB91,G97,O02,HNG04,HG11}.
Theoretical approaches can be found in references
\cite{LGZJ77,MF83,MF84,D86,SFLD92,D89}. Recent results in reference 
\cite{DG20} make various predictions in models of confined star polymers.  

\begin{figure}[h]
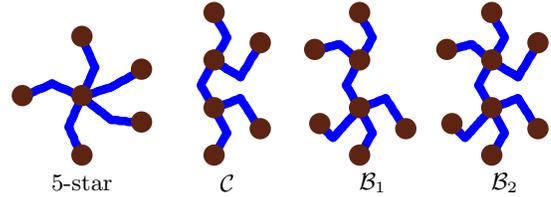

\beginpicture
\setcoordinatesystem units <0.6pt,0.6pt>
\setplotarea x from -90 to 150, y from 0 to 110
\setplotarea x from -90 to 250, y from 0 to 90

\setcoordinatesystem units <0.5pt,0.5pt> point at 0 0 
\setplotsymbol ({\scriptsize$\bullet$})
\color{Blue}
\plot 0 0 -10 22.5 0 45 /
\plot 0 45 10 67.5 0 90 /
\plot 0 45 22.5 27.5 45 25 /
\plot 0 45 22.5 52.5 45 65 /
\plot 0 45 -22.5 55 -45 45 /
\color{Sepia}
\multiput {\huge$\bullet$} at 0 0 0 45 0 90 45 25 45 65 -45 45 /
\color{Black}
\put {5-star} at 0 -20 

\setcoordinatesystem units <0.5pt,0.5pt> point at -30 0 
\setplotsymbol ({\scriptsize$\bullet$})
\color{Blue}
\plot 70 0 80 18 70 36 60 54 70 72 80 90 70 108 /
\plot 70 36 90 44 105 20 /
\plot 70 72 90 60 105 85 /
\color{Sepia}
\multiput {\huge$\bullet$} at 70 0 70 36 70 72 70 108 105 20 105 85 /
\color{Black}
\put {$\C{C}$} at 80 -20 

\setcoordinatesystem units <0.5pt,0.5pt> point at -140 0 
\setplotsymbol ({\scriptsize$\bullet$})
\color{Blue}
\plot 70 0 80 18 70 36 60 54 70 72 80 90 70 108 /
\plot 70 36 90 44 105 20 /
\plot 70 36 50 14 40 24 /
\plot 70 72 55 85 36 80 /
\color{Sepia}
\multiput {\huge$\bullet$} at 70 0 70 36 70 72 70 108 105 20 40 24 36 80 /
\color{Black}
\put {$\C{B}_1$} at 80 -20 

\setcoordinatesystem units <0.5pt,0.5pt> point at -240 0 
\setplotsymbol ({\scriptsize$\bullet$})
\color{Blue}
\plot 70 0 80 18 70 36 60 54 70 72 80 90 70 108 /
\plot 70 36 90 44 105 20 /
\plot 70 36 50 14 40 24 /
\plot 70 72 90 60 105 85 /
\plot 70 72 55 85 36 80 /
\color{Sepia}
\multiput {\huge$\bullet$} at 70 0 70 36 70 72 70 108 
105 20 40 24 105 85 36 80 /
\color{Black}
\put {$\C{B}_2$} at 80 -20 

\normalcolor
\setlinear
\endpicture
\caption{From the left, schematic diagrams of the connectivity of a
$5$-star, a comb $\C{C}$, a brush $\C{B}_1$, and
a brush $\C{B}_2$.}
\label{f1}
\end{figure}

In this paper we use a parallel implementation of the flatGARM algorithm 
\cite{CJvR20} to estimate two dimensional values of the star vertex
exponents.  In addition, we use a parallel implementation of the Wang-Landau
algorithm \cite{WL01,Z05,Z08,Z08A} to estimate the entropic exponents 
of monodispersed acyclic branched networks in the square lattice. In
particular, we test the scaling relation \cite{D86}
\begin{equation}
\gamma_{\C{G}}-1 = \sum_{f\geq 1} m_f \, \sigma_f
\label{1}
\end{equation}
where $\C{G}$ is the connectivity of the branched network (see figure
\ref{f1}).  This relates the entropic exponent $\gamma_{\C{G}}$ of
the network to the star vertex exponents $\sigma_f$.   

In our implementation of flatGARM we sampled square lattice $f$-stars
to lengths $1000$ steps (edges) per arm, for $f\in\{3,4,5,6\}$.  Monodispersed
branch networks (a comb and two brushes) with underlying connectivity
shown in figure \ref{f1} were sampled using the Wang-Landau algorithm
to lengths of $200$ steps per branch.

The branches (arms) of lattice stars and monodispersed branched
networks (figure \ref{f1}) are self-avoiding walks (also avoiding each other)
joining nodes of various degrees.  In an $f$-star, the branches join the central 
node of degree $f$ to $f$ nodes of degree $1$, and the total length of
the star is $f n$, if each branch has length $n$.  Similarly, the
length of a uniform network consisting of $b$ branches each of length $n$
is $b\thin n$, and the branches are self-avoiding walks joining 
the nodes in the network.

A lattice $f$-star is \textit{almost uniform} if the length of the longest
arms exceed the length of the shortest arms by exactly one.  If the arms have
the same length, then it is \textit{uniform}.  A lattice star will be 
\textit{monodispersed} if it is uniform, or almost uniform.  Uniform,
almost uniform and monodispersed branched networks are similarly defined.

Denote by $s_n^{(f)}$ the number of monodispersed lattice stars 
of total length $n$, and with $f$ arms.  The \textit{growth constant} 
$\mu_2$ is defined by
\begin{equation}
\lim_{n\to\infty} \Sfrac{1}{n} \log s_n^{(f)} = \log \mu_2 .
\label{2}
\end{equation}
This limit is known to exist in $d$-dimensions \cite{WLWG86,SSW88A,SW89,WS92,SS93}
for uniform $f$-stars (that is, if $n=fm$ as $m\to\infty$), and $\mu_d$ is 
equal to the growth constant of self-avoiding walks.   The methods in 
references \cite{WLWG86,SSW88A} can also be used to prove 
this for monodisperse lattice $f$-stars.  We classify monodisperse lattice $f$-stars 
of length $n=fm{+}k$ according to the remainder $k\in\{0,1,2,\ldots,f{-}1\}$.  Uniform
stars are in the class $k=0$ while almost uniform stars are in the classes with
$1\leq k < f$.

Denote by $c_n$ the number of self-avoiding walks from the origin.  
There is substantial numerical evidence that
\begin{equation}
c_n = C\, n^{\gamma-1}\,\mu_2^n\,(1+o(1))
\label{3}
\end{equation}
where $\gamma$ is the entropic exponent.  In two dimensions
$\gamma=43/32$ is exact \cite{N82,N87}. In analogy with equation \Ref{3}
the asymptotic behaviour of $s_n^{(f)}$ is   
\begin{eqnarray}
s_{mf+k}^{(f)} = C^{(f)}_k\, n^{\gamma_f-1}\,\mu_2^n\,(1+o(1)) ,
\label{4}
\end{eqnarray}
in the square lattice,
where $k$ is fixed in $\{0,1,2,\ldots,f{-}1\}$ and where $n=fm{+}k$,
and $\mu_2$ is the self-avoiding square lattice growth constant.
Only the amplitude $C^{(f)}_k$ is dependent on the class
of monodispersed stars, while the entropic exponent $\gamma_f$ is
dependent only on the number of arms.  Parity effects in $s_n^{(f)}$
(due to both the lattice, and the number of arms $f$) are present in 
the $o(1)$ correction term, and so decay with increasing $n$.

Equations \Ref{2} and \Ref{4} can be generalised to square lattice stars with
$f>4$ arms by using more than one central node as shown in figure \ref{f2}.   
The edge joining the two central nodes does not
count towards the total length of the star.

\def\stepb#1#2#3#4{\color{Blue}
                                   \plot #1 #2 #3 #4  /
                                   \color{black} 
                                   \multiput {\large$\bullet$} at #1 #2 #3 #4 /
                                   }
\def\stepr#1#2#3#4{\color{Red}
                                   \plot #1 #2 #3 #4  /
                                   \color{black} 
                                   \multiput {\large$\bullet$} at #1 #2 #3 #4 /
                                   }
\def\stepg#1#2#3#4{\color{DarkGreen}
                                   \plot #1 #2 #3 #4  /
                                   \color{black} 
                                   \multiput {\large$\bullet$} at #1 #2 #3 #4 /
                                   }
\def\putnum#1#2#3{\put {\scalebox{0.65}{#1}} at #2 #3 }

\begin{figure}[h]
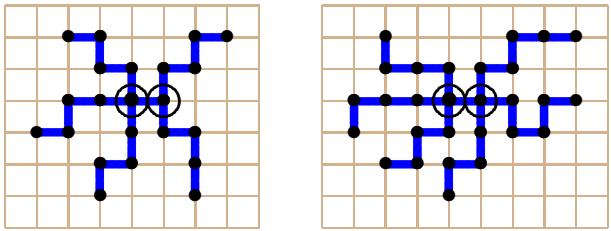

\beginpicture
\setcoordinatesystem units <1.2pt,1.2pt> point at 0 0 
\setplotarea x from -50 to 30, y from -40 to 30
\setplotarea x from -40 to 40, y from -40 to 30

\color{Tan}
\grid 8 7

\setplotsymbol ({\scriptsize$\bullet$})

\stepb{0}{0}{10}{0}
\stepb{0}{0}{-10}{0}
\stepb{0}{0}{0}{10}
\stepb{0}{0}{0}{-10}
\stepb{10}{0}{10}{10}
\stepb{10}{0}{10}{-10}

\stepb{-10}{0}{-20}{0}
\stepb{-20}{0}{-20}{-10}
\stepb{-20}{-10}{-30}{-10}
\stepb{0}{-10}{0}{-20}
\stepb{0}{-20}{-10}{-20}
\stepb{-10}{-20}{-10}{-30}
\stepb{0}{10}{-10}{10}
\stepb{-10}{10}{-10}{20}
\stepb{-10}{20}{-20}{20}
\stepb{10}{-10}{20}{-10}
\stepb{20}{-10}{20}{-20}
\stepb{20}{-20}{20}{-30}
\stepb{10}{10}{20}{10}
\stepb{20}{10}{20}{20}
\stepb{20}{20}{30}{20}

\color{black}
\put {\Large$\bullet$} at 0 0 
\setplotsymbol ({.})
\circulararc 360 degrees from 5 0 center at 0 0
\circulararc 360 degrees from 15 0 center at 10 0

\setcoordinatesystem units <1.2pt,1.2pt> point at -100 0 
\setplotarea x from -40 to 50, y from -40 to 30

\color{Tan}
\grid 9 7 

\setplotsymbol ({\scriptsize$\bullet$})

\stepb{0}{0}{10}{0}
\stepb{0}{0}{-10}{0}
\stepb{0}{0}{0}{10}
\stepb{0}{0}{0}{-10}

\stepb{-10}{0}{-20}{0}
\stepb{-20}{0}{-30}{0}
\stepb{-30}{0}{-30}{-10}

\stepb{0}{-10}{-10}{-10}
\stepb{-10}{-10}{-10}{-20}
\stepb{-10}{-20}{-20}{-20}

\stepb{0}{10}{-10}{10}
\stepb{-10}{10}{-20}{10}
\stepb{-20}{10}{-20}{20}

\stepb{10}{0}{10}{-10}
\stepb{10}{-10}{10}{-20}
\stepb{10}{-20}{0}{-20}
\stepb{0}{-20}{0}{-30}

\stepb{10}{0}{20}{0}
\stepb{20}{0}{20}{-10}
\stepb{20}{-10}{30}{-10}
\stepb{30}{-10}{30}{0}
\stepb{30}{0}{40}{0}

\stepb{10}{0}{10}{10}
\stepb{10}{10}{20}{10}
\stepb{20}{10}{20}{20}
\stepb{20}{20}{30}{20}
\stepb{30}{20}{40}{20}

\color{black}
\put {\Large$\bullet$} at 0 0 
\setplotsymbol ({.})
\circulararc 360 degrees from 5 0 center at 0 0
\circulararc 360 degrees from 15 0 center at 10 0

\normalcolor

\endpicture
\caption{A uniform square lattice $5$-star and a $6$-star.
There are two central nodes accommodating the arms.  
Since the edge joining the two central nodes is not counted as 
part of the length of the star, then $5$-star on the left has 
length $20$, and the $6$-star on the right has length $24$.}
\label{f2}
\end{figure}

The entropic exponent $\gamma_f$ of $f$-stars is related to 
\textit{vertex exponents} $\sigma_f$ by \cite{D86,D89}
\begin{equation}
\gamma_f-1 = \sigma_f + f\,\sigma_1 .
\label{5} 
\end{equation}
More generally, the vertex exponents are associated with nodes
in stars and more general monodispersed branched networks:
$\sigma_1$ is associated with end-vertices of degree $1$ (end-points of 
branches), while the $\sigma_f$ with $f\geq3$ are associated with nodes of
degree $f$ in the networks.  If $f=1$, then $\gamma_1$ is the
entropic exponent $\gamma$ of self-avoiding walks, with exact value
$\gamma_1= 43/32$ in two dimensions \cite{N82,N87}.  By
equation \Ref{5}, $\sigma_1=11/64$.  If $f=2$, then the star has two
arms and so is a self-avoiding walk.  This shows by equation \Ref{5} that
$\sigma_2=0$.  Exact values for the other vertex exponents
are similarly calculated from equation \Ref{5} and are given by
\cite{D86,N82,N87}
\begin{equation}
\sigma_f = \Sfrac{1}{16}+\Sfrac{1}{4}\,f-\Sfrac{9}{64}\,f^2.
\label{6} 
\end{equation}
We show the exact values and estimates of $\sigma_f$ in two dimensions
for $f\leq 6$ in table \ref{t1}, and compare it to the results in reference
\cite{WGLW86}, and with the values obtained in this paper.
Our results confirm to within numerical accuracy the exact values.

\def\0{\phantom{-}}
\def\1{\phantom{0}}
\begin{table}[h!]
\caption{Vertex exponents in 2 dimensions}
\setlength{\extrarowheight}{1.7pt}
\setlength{\tabcolsep}{8pt}
\begin{tabular} {llcr}           
\hline   
$f$& \1\1Exact & \cite{WGLW86}  & This work\scalebox{2.3}{$ $} \cr 
\hline
$\sigma_1$ & $\00.171875$ & $\;-$ & $\,{0.17188(12)}$  \cr
$\sigma_3$ & $-0.453125$     &  $-0.45(2)$& ${-0.45282(69)}$  \cr
$\sigma_4$ & $-1.1875$     & $-1.17(4)$ & ${-1.1864(27)\1}$  \cr
$\sigma_5$ & $-2.203125$   & $-2.14(4)$ & ${-2.2016(19)\1}$  \cr
$\sigma_6$ & $-3.5$ & $-3.36(5)$ & ${-3.4981(27)\1}$  \cr
\hline
\end{tabular}
\label{t1} 
\end{table}

Lattice networks are  \textit{uniform} if all their branches are 
self-avoiding walks of the same length $m$.  If a uniform lattice network 
of connectivity $\C{G}$ has $b$ branches and $n=b\thin m$ edges, then the total 
number of such networks (up to equivalency under translations) is denoted by 
$c_{n}(\C{G})$.  It is generally accepted that
\begin{equation}
c_{n}(\C{G}) = C_\C{G}\, n^{\gamma_\C{G} - 1}\, \mu_2^{n}\,(1+o(1)) ,
\label{7} 
\end{equation}
where the growth constant is equal to that of square lattice self-avoiding walks 
\cite{SS93,SW89,SSW88A,WS92}.   The relation of $\gamma_\C{G}$ 
to the vertex exponents is given by
\begin{equation}
\gamma_\C{G} - 1 = \sum_{f\geq 1} m_f \, \sigma_f - c(\C{G}) \, d\nu,
\label{8} 
\end{equation}
where $m_f$ is the number of vertices of degree $f$, and where $c(\C{G})$ is
the cyclomatic index (the number of independent cycles) in the network
\cite{D86,D89}.  The networks in figure \ref{1} are acyclic, and by the above 
\begin{eqnarray}
\gamma_{\C{C}} - 1 & = &  4\,\sigma_1 + 2\, \sigma_3 , \nonumber \\ 
\gamma_{\C{B}_1} - 1 & = & 5\,\sigma_1 + \sigma_3 + \sigma_4,
\label{9}  \\
\gamma_{\C{B}_2} - 1 & = & 6\,\sigma_1 + 2\, \sigma_4 . \nonumber 
\end{eqnarray}
The exact values of the $\gamma_{\C{G}}$ are obtained from these
relations assuming that the scaling relation in equation \Ref{8} holds, and
are listed in the second column of table \ref{t2}.  In the third column we
list estimates obtained using equation \Ref{9} and the numerical
estimates listed in table \ref{t1}, and in the last column the direct
estimates from our Wang-Landau data for lattice networks.  These 
results show excellent agreement with both the exact values and the 
numerical data using equation \Ref{9} which is strong numerical 
evidence that equation \Ref{9} applies to the branched polymer 
networks shown in figure \ref{f1}.

\begin{table}[t]
\caption{$\gamma_{\mathcal{G}}{-}1$ for lattice networks}
\setlength{\extrarowheight}{1.7pt}
\setlength{\tabcolsep}{8pt}
\begin{tabular}{llcc}
\hline                          
$\C{G}$ & \1\1Exact & Eqn \Ref{9} & This work$\0$  \cr 
\hline
$\C{C}$ & $-0.21875$  & $-0.2181(14)$ & $-0.2187(22)$  \cr
$\C{B}_1$ & $-0.78125$ & $-0.7799(34)$ & $-0.7817(40)$  \cr
$\C{B}_2$ & $-1.34375$ & $-1.3412(54)$ & $-1.3426(82)$  \cr
\hline
\end{tabular}
\label{t2}
\end{table}

\section{Numerical simulations}
\label{S2}  

\subsection{Determining $\sigma_1$}

The numbers $c_n$ in equation \Ref{3} were estimated by sampling
self-avoiding walks to length $10,\!000$ with the parallel flatPERM algorithm 
\cite{G97,PK04,HG11,CJvR20} ($12$ parallel sequences for a total of 
$2.65\times 10^9$ iterations).  In two dimensions $\gamma = 1.34375$
and the $o(1)$ term in equation \Ref{3} is believed to be a powerlaw correction 
of the form $A\,n^{-1} + B\, n^{-\Delta}$ (where $\Delta = 3/2$ 
in two dimensions \cite{CGJPRS05} and is the leading confluent correction
exponent).

\begin{table}[t!]
\caption{Least squares fits of $c_n$ in the square lattice}
\setlength{\extrarowheight}{1.7pt}
\begin{tabular}{lc}        
\hline                
$n_{min}$& 
$\1\1\1 n\, \log \LB c_n / c_{n-2} \RB
\simeq 2\,(\gamma-1) + 2n\, \log \mu_2 + A / n$   \cr 
\hline
$10$ & $0.6870990 + 1.94016321\, n  -0.16886256\, / n$ \cr
 $20$ & $0.6870978 + 1.94016321\, n  -0.16749947\, / n$ \cr
 $30$ & $0.6870860 + 1.94016321\, n  -0.15795150\, / n$ \cr
 $40$ & $0.6870813 + 1.94016321\, n  -0.15382565\, / n$ \cr
 $50$ & $0.6870635 + 1.94016322\, n  -0.13736803\, / n$ \cr
\hline
\end{tabular}
\label{t3}
\end{table}

We use the ratio $c_n/c_{n-2}$ and the model 
\begin{equation}
n\, \log \LB \frac{c_n}{c_{n-2}} \RB
\simeq 2\,(\gamma-1) + 2n\, \log \mu_2 + A\, n^{-\Delta_1}  
\label{10}
\end{equation}
to estimate $\gamma$.  Linear least-squares fits (with 
$\Delta_1 \in \{0.5,1.0,1.5\}$) were done for $n$ greater than or 
equal to $n_{min}$ where $n_{min}\in\{10,20,\ldots,100\}$.  
The results for $\Delta_1=1$ and $n_{min} \leq 50$ are shown in 
table \ref{t3}. For each value of $\Delta_1$ the results were extrapolated
against $n_{min}$ using the model $c_0 + c_1\,/n_{min} + c_2\,/n^2_{min}$
and  comparing the results for the choices of $\Delta_1$, the estimate
\begin{equation}
\gamma = 1.34359(23) 
\label{11}
\end{equation}
was obtained (the error bar is the largest difference between the
average and the estimates).  Since $\sigma_1  = (\gamma{-}1)/2$, this
gives
\begin{equation}
\sigma_1 = 0.17188(12) ,
\label{12}
\end{equation} 
consistent within its error bar with the exact value $\sigma_1=0.171875$.

\subsection{Calculating $\sigma_f$ for $3\leq f\leq 6$}

An $f$-star is grown by the GARM algorithm \cite{RJvR08} by adding
steps to the endpoints of the arms in a cyclic order.  The algorithm
is an approximate enumeration algorithm, and it estimates numbers
$u_n^{(f)}$ of $f$-stars of length $n$.  To relate $u_n^{(f)}$ to
$s_n^{(f)}$ equation \Ref{4}, first note that the algorithm imposes
ordering of the arms:  it adds a step to the first arm, then the second
arm, and so on.  This shows that $u_n^{(f)}$ is the number of $f$-stars
with \textit{labelled arms} (while $s_n^{(f)}$ is the number of $f$-stars
with unlabelled arms).  To determine the symmetry factor relating $s_n^{(f)}$
and $u_n^{(f)}$, note that a monodisperse $f$-star of length $n$ has 
$k$ arms of length $\lcl n/f\rcl$ and $f{-}k$ arms of length $\lfl n/f \rcl$.  
Since the $k$ longest arms can be ordered in $k!$ ways, 
and the $f{-}k$ shortest arms in $(f{-}k)!$ ways, a symmetry factor
of $k!\,(f{-}k)!$ is introduced. This is particularly true for $3$-stars and
$4$-stars in the cubic lattice, so if $n=mf{+}k$, then by equation \Ref{4} 
the algorithm estimates the numbers
\begin{equation}
u_n^{(f)} 
= k!\,(f{-}k)!\; C_k^{(f)}\, n^{\gamma_f-1}\,\mu_2^n\,(1+o(1)) .
\label{13}
\end{equation}
for $f=3$ or $f=4$ in the square lattice.  The $o(1)$ correction contains,
in addition to analytic and confluent correction terms, parity effects due
to the lattice and the number of arms.  Our results show that the parity
effects decay quickly with increasing $n$.

Similar arguments for $5$- and $6$-stars give 
\begin{equation}
u_n^{(f)} = V_k^{(f)}\,s_n^{(f)}
\label{13a}
\end{equation}
where $n=m\,f{+}k$ (excluding the extra edge between the two central 
nodes), and where the symmetry factor is given by
\begin{equation}
V_k^{(f)} = \cases{
\lfl f/2 \rfl\,!\,(3-k)!\,k!, & \hbox{if $0 \leq k < 3$}; \cr
3!\,(f-k)!\,(k-3)!, & \hbox{if $3 \leq k \leq f{-}1$}. }
\label{14}   
\end{equation}

\subsubsection{Estimating $\sigma_f$ numerically for $3\leq f \leq 6$}

Square lattice $f$-stars for $3\leq f \leq 6$ were sampled a total of $4\times 10^9$ 
started flatGARM \cite{RJvR08} sequences along $4$ parallel threads for lengths 
up to $1,\!000$ steps per branch (arm).  These simulations produced estimates
of $u_n^{(f)}$ in equations \Ref{13} and \Ref{13a}. 

To estimate $\gamma_f{-}1$ from our data, notice that if 
$x=\gamma_f{-}1$, then
\begin{equation}
Q_n(x) = \log \LB \frac{u_n^{(f)}}{\mu_d^n\, n^x} \RB \simeq C_0 + C_1 n^{-1},
\label{15}
\end{equation}
where $x=\sigma_f{+}f\sigma_1$.  By using the best estimate of $\mu_2$ 
in the literature ($\mu_2=2.63815853035(2)$ \cite{CJ12}), we determine 
that value of $x$ so that $Q_n(x)$ approaches a constant as $n$ increases.

\begin{figure}[t!]
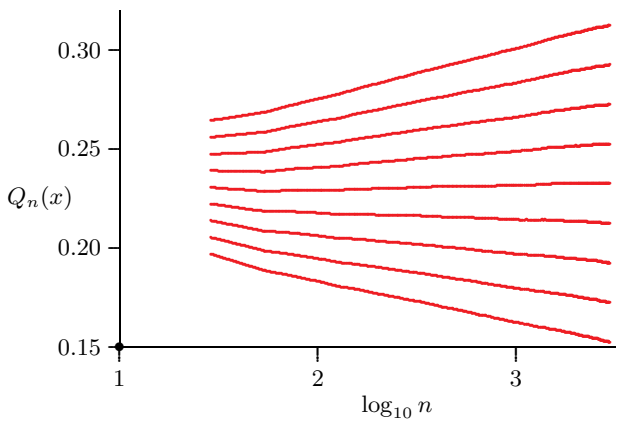

\normalcolor
\color{black}
\beginpicture
\setcoordinatesystem units <75pt,750pt>
\setplotarea x from 1 to 3.5, y from 0.15 to 0.32

\setplotsymbol ({\tiny.})
\axis left shiftedto x=1 
 /
\axis bottom shiftedto y=0.15
/

\plot 1 0.15 0.95 0.15 /  \put {$0.15$} at 0.80 0.15
\plot 1 0.20 0.95 0.20 /  \put {$0.20$} at 0.80 0.20
\plot 1 0.25 0.95 0.25 /  \put {$0.25$} at 0.80 0.25
\plot 1 0.30 0.95 0.30 /  \put {$0.30$} at 0.80 0.30

\plot 1 0.15 1 0.143 /  \put {$1$} at 1 0.134
\plot 2 0.15 2 0.143 /  \put {$2$} at 2 0.134
\plot 3 0.15 3 0.143 /  \put {$3$} at 3 0.134

\put {\footnotesize$\bullet$} at 1 0.15

\put {$\log_{10} n$} at 2.4 0.12
\put {$Q_n(x)$} at 0.6 0.225

\setplotsymbol ({\scalebox{0.3}{$\bullet$}})
\color{Red}
\plot 
1.462 0.2642  1.732 0.2683  1.898 0.2727  2.017 0.2755  2.111 0.2776  2.188 0.2799  2.253 0.2817  2.310 0.2830  2.360 0.2844  2.405 0.2856  2.446 0.2866  2.483 0.2877  2.517 0.2886  2.549 0.2894  2.579 0.2901  2.606 0.2908  2.632 0.2914  2.657 0.2920  2.680 0.2926  2.702 0.2932  2.723 0.2938  2.744 0.2942  2.763 0.2947  2.781 0.2952  2.799 0.2956  2.816 0.2959  2.832 0.2963  2.848 0.2968  2.863 0.2971  2.877 0.2975  2.892 0.2978  2.905 0.2982  2.919 0.2985  2.931 0.2989  2.944 0.2991  2.956 0.2994  2.968 0.2997  2.980 0.2999  2.991 0.3002  3.002 0.3005  3.012 0.3008  3.023 0.3010  3.033 0.3013  3.043 0.3016  3.053 0.3019  3.062 0.3022  3.072 0.3024  3.081 0.3026  3.090 0.3028  3.098 0.3030  3.107 0.3033  3.115 0.3036  3.124 0.3039  3.132 0.3042  3.140 0.3044  3.147 0.3047  3.155 0.3048  3.163 0.3050  3.170 0.3051  3.177 0.3053  3.184 0.3056  3.191 0.3058  3.198 0.3060  3.205 0.3061  3.212 0.3063  3.219 0.3065  3.225 0.3067  3.231 0.3068  3.238 0.3069  3.244 0.3070  3.250 0.3072  3.256 0.3073  3.262 0.3075  3.268 0.3076  3.274 0.3078  3.280 0.3080  3.285 0.3082  3.291 0.3083  3.296 0.3085  3.302 0.3086  3.307 0.3087  3.313 0.3087  3.318 0.3088  3.323 0.3089  3.328 0.3090  3.333 0.3092  3.338 0.3093  3.343 0.3094  3.348 0.3096  3.353 0.3097  3.358 0.3098  3.362 0.3099  3.367 0.3100  3.372 0.3101  3.376 0.3101  3.381 0.3103  3.385 0.3104  3.390 0.3105  3.394 0.3106  3.399 0.3108  3.403 0.3109  3.407 0.3111  3.411 0.3112  3.416 0.3112  3.420 0.3113  3.424 0.3113  3.428 0.3114  3.432 0.3114  3.436 0.3114  3.440 0.3115  3.444 0.3116  3.448 0.3117  3.452 0.3118  3.455 0.3119  3.459 0.3119  3.463 0.3120  3.467 0.3122  3.470 0.3123  3.474 0.3124 
 /
 
 \plot 1.462 0.2558  1.732 0.2583  1.898 0.2618  2.017 0.2639  2.111 0.2654  2.188 0.2674  2.253 0.2687  2.310 0.2697  2.360 0.2708  2.405 0.2718  2.446 0.2726  2.483 0.2734  2.517 0.2741  2.549 0.2747  2.579 0.2753  2.606 0.2758  2.632 0.2762  2.657 0.2767  2.680 0.2772  2.702 0.2776  2.723 0.2781  2.744 0.2784  2.763 0.2788  2.781 0.2792  2.799 0.2795  2.816 0.2797  2.832 0.2800  2.848 0.2804  2.863 0.2807  2.877 0.2809  2.892 0.2811  2.905 0.2814  2.919 0.2817  2.931 0.2820  2.944 0.2822  2.956 0.2824  2.968 0.2826  2.980 0.2828  2.991 0.2830  3.002 0.2832  3.012 0.2834  3.023 0.2836  3.033 0.2839  3.043 0.2841  3.053 0.2843  3.062 0.2845  3.072 0.2847  3.081 0.2849  3.090 0.2850  3.098 0.2852  3.107 0.2854  3.115 0.2857  3.124 0.2859  3.132 0.2862  3.140 0.2863  3.147 0.2865  3.155 0.2867  3.163 0.2868  3.170 0.2869  3.177 0.2871  3.184 0.2873  3.191 0.2874  3.198 0.2876  3.205 0.2877  3.212 0.2878  3.219 0.2880  3.225 0.2881  3.231 0.2882  3.238 0.2882  3.244 0.2883  3.250 0.2885  3.256 0.2886  3.262 0.2887  3.268 0.2888  3.274 0.2890  3.280 0.2892  3.285 0.2893  3.291 0.2894  3.296 0.2895  3.302 0.2896  3.307 0.2896  3.313 0.2897  3.318 0.2897  3.323 0.2897  3.328 0.2899  3.333 0.2900  3.338 0.2901  3.343 0.2902  3.348 0.2903  3.353 0.2904  3.358 0.2905  3.362 0.2906  3.367 0.2907  3.372 0.2907  3.376 0.2907  3.381 0.2908  3.385 0.2909  3.390 0.2910  3.394 0.2911  3.399 0.2912  3.403 0.2914  3.407 0.2915  3.411 0.2915  3.416 0.2915  3.420 0.2916  3.424 0.2916  3.428 0.2916  3.432 0.2916  3.436 0.2917  3.440 0.2917  3.444 0.2918  3.448 0.2919  3.452 0.2919  3.455 0.2920  3.459 0.2920  3.463 0.2921  3.467 0.2922  3.470 0.2923  3.474 0.2924  
/

\plot 1.462 0.2473  1.732 0.2483  1.898 0.2509  2.017 0.2523  2.111 0.2533  2.188 0.2548  2.253 0.2557  2.310 0.2564  2.360 0.2572  2.405 0.2579  2.446 0.2585  2.483 0.2591  2.517 0.2596  2.549 0.2600  2.579 0.2605  2.606 0.2608  2.632 0.2611  2.657 0.2614  2.680 0.2618  2.702 0.2621  2.723 0.2624  2.744 0.2626  2.763 0.2629  2.781 0.2632  2.799 0.2634  2.816 0.2635  2.832 0.2637  2.848 0.2640  2.863 0.2642  2.877 0.2644  2.892 0.2645  2.905 0.2647  2.919 0.2649  2.931 0.2651  2.944 0.2652  2.956 0.2653  2.968 0.2655  2.980 0.2656  2.991 0.2657  3.002 0.2659  3.012 0.2661  3.023 0.2662  3.033 0.2664  3.043 0.2666  3.053 0.2668  3.062 0.2669  3.072 0.2670  3.081 0.2672  3.090 0.2673  3.098 0.2674  3.107 0.2675  3.115 0.2677  3.124 0.2679  3.132 0.2681  3.140 0.2683  3.147 0.2684  3.155 0.2685  3.163 0.2686  3.170 0.2686  3.177 0.2688  3.184 0.2689  3.191 0.2691  3.198 0.2692  3.205 0.2692  3.212 0.2693  3.219 0.2694  3.225 0.2696  3.231 0.2696  3.238 0.2696  3.244 0.2696  3.250 0.2698  3.256 0.2698  3.262 0.2699  3.268 0.2700  3.274 0.2701  3.280 0.2703  3.285 0.2704  3.291 0.2705  3.296 0.2705  3.302 0.2706  3.307 0.2706  3.313 0.2706  3.318 0.2706  3.323 0.2706  3.328 0.2707  3.333 0.2708  3.338 0.2708  3.343 0.2710  3.348 0.2710  3.353 0.2711  3.358 0.2712  3.362 0.2712  3.367 0.2713  3.372 0.2713  3.376 0.2713  3.381 0.2714  3.385 0.2714  3.390 0.2715  3.394 0.2715  3.399 0.2716  3.403 0.2718  3.407 0.2719  3.411 0.2719  3.416 0.2719  3.420 0.2719  3.424 0.2719  3.428 0.2719  3.432 0.2719  3.436 0.2719  3.440 0.2719  3.444 0.2720  3.448 0.2720  3.452 0.2720  3.455 0.2721  3.459 0.2721  3.463 0.2722  3.467 0.2723  3.470 0.2723  3.474 0.2724 
/
 
 \plot 1.462 0.2389  1.732 0.2383  1.898 0.2399  2.017 0.2407  2.111 0.2411  2.188 0.2422  2.253 0.2428  2.310 0.2432  2.360 0.2436  2.405 0.2441  2.446 0.2444  2.483 0.2449  2.517 0.2451  2.549 0.2454  2.579 0.2456  2.606 0.2458  2.632 0.2459  2.657 0.2461  2.680 0.2464  2.702 0.2465  2.723 0.2467  2.744 0.2468  2.763 0.2470  2.781 0.2471  2.799 0.2473  2.816 0.2473  2.832 0.2474  2.848 0.2476  2.863 0.2477  2.877 0.2478  2.892 0.2478  2.905 0.2480  2.919 0.2481  2.931 0.2482  2.944 0.2483  2.956 0.2483  2.968 0.2484  2.980 0.2485  2.991 0.2485  3.002 0.2486  3.012 0.2488  3.023 0.2488  3.033 0.2490  3.043 0.2491  3.053 0.2492  3.062 0.2493  3.072 0.2493  3.081 0.2494  3.090 0.2495  3.098 0.2495  3.107 0.2497  3.115 0.2498  3.124 0.2499  3.132 0.2501  3.140 0.2502  3.147 0.2503  3.155 0.2503  3.163 0.2504  3.170 0.2504  3.177 0.2505  3.184 0.2506  3.191 0.2507  3.198 0.2508  3.205 0.2508  3.212 0.2508  3.219 0.2509  3.225 0.2510  3.231 0.2510  3.238 0.2510  3.244 0.2510  3.250 0.2511  3.256 0.2511  3.262 0.2511  3.268 0.2512  3.274 0.2513  3.280 0.2514  3.285 0.2515  3.291 0.2515  3.296 0.2515  3.302 0.2516  3.307 0.2515  3.313 0.2515  3.318 0.2515  3.323 0.2515  3.328 0.2516  3.333 0.2516  3.338 0.2516  3.343 0.2517  3.348 0.2518  3.353 0.2518  3.358 0.2518  3.362 0.2519  3.367 0.2519  3.372 0.2518  3.376 0.2518  3.381 0.2519  3.385 0.2519  3.390 0.2520  3.394 0.2520  3.399 0.2521  3.403 0.2522  3.407 0.2522  3.411 0.2522  3.416 0.2522  3.420 0.2522  3.424 0.2522  3.428 0.2522  3.432 0.2521  3.436 0.2521  3.440 0.2521  3.444 0.2521  3.448 0.2522  3.452 0.2522  3.455 0.2522  3.459 0.2522  3.463 0.2522  3.467 0.2523  3.470 0.2523  3.474 0.2524 
/

\plot 1.462 0.2305  1.732 0.2284  1.898 0.2290  2.017 0.2291  2.111 0.2290  2.188 0.2296  2.253 0.2298  2.310 0.2299  2.360 0.2301  2.405 0.2303  2.446 0.2303  2.483 0.2306  2.517 0.2306  2.549 0.2307  2.579 0.2308  2.606 0.2308  2.632 0.2308  2.657 0.2308  2.680 0.2309  2.702 0.2310  2.723 0.2310  2.744 0.2310  2.763 0.2311  2.781 0.2311  2.799 0.2311  2.816 0.2311  2.832 0.2311  2.848 0.2312  2.863 0.2312  2.877 0.2312  2.892 0.2312  2.905 0.2313  2.919 0.2313  2.931 0.2314  2.944 0.2313  2.956 0.2313  2.968 0.2313  2.980 0.2313  2.991 0.2313  3.002 0.2314  3.012 0.2314  3.023 0.2314  3.033 0.2315  3.043 0.2316  3.053 0.2316  3.062 0.2317  3.072 0.2317  3.081 0.2317  3.090 0.2317  3.098 0.2317  3.107 0.2318  3.115 0.2319  3.124 0.2320  3.132 0.2321  3.140 0.2321  3.147 0.2322  3.155 0.2322  3.163 0.2322  3.170 0.2321  3.177 0.2322  3.184 0.2323  3.191 0.2323  3.198 0.2324  3.205 0.2323  3.212 0.2324  3.219 0.2324  3.225 0.2324  3.231 0.2324  3.238 0.2323  3.244 0.2323  3.250 0.2323  3.256 0.2324  3.262 0.2324  3.268 0.2324  3.274 0.2325  3.280 0.2325  3.285 0.2326  3.291 0.2326  3.296 0.2326  3.302 0.2325  3.307 0.2325  3.313 0.2325  3.318 0.2324  3.323 0.2324  3.328 0.2324  3.333 0.2324  3.338 0.2324  3.343 0.2325  3.348 0.2325  3.353 0.2325  3.358 0.2325  3.362 0.2325  3.367 0.2325  3.372 0.2324  3.376 0.2324  3.381 0.2324  3.385 0.2324  3.390 0.2324  3.394 0.2325  3.399 0.2325  3.403 0.2326  3.407 0.2326  3.411 0.2326  3.416 0.2325  3.420 0.2325  3.424 0.2325  3.428 0.2324  3.432 0.2324  3.436 0.2323  3.440 0.2323  3.444 0.2323  3.448 0.2324  3.452 0.2323  3.455 0.2323  3.459 0.2323  3.463 0.2323  3.467 0.2324  3.470 0.2324  3.474 0.2324
/

\plot 1.462 0.2221  1.732 0.2184  1.898 0.2181  2.017 0.2175  2.111 0.2168  2.188 0.2170  2.253 0.2168  2.310 0.2166  2.360 0.2165  2.405 0.2164  2.446 0.2163  2.483 0.2163  2.517 0.2162  2.549 0.2160  2.579 0.2159  2.606 0.2158  2.632 0.2156  2.657 0.2156  2.680 0.2155  2.702 0.2154  2.723 0.2154  2.744 0.2153  2.763 0.2152  2.781 0.2151  2.799 0.2150  2.816 0.2149  2.832 0.2148  2.848 0.2148  2.863 0.2148  2.877 0.2147  2.892 0.2146  2.905 0.2146  2.919 0.2145  2.931 0.2145  2.944 0.2144  2.956 0.2143  2.968 0.2143  2.980 0.2141  2.991 0.2141  3.002 0.2141  3.012 0.2141  3.023 0.2140  3.033 0.2140  3.043 0.2140  3.053 0.2141  3.062 0.2140  3.072 0.2140  3.081 0.2140  3.090 0.2139  3.098 0.2139  3.107 0.2139  3.115 0.2139  3.124 0.2140  3.132 0.2141  3.140 0.2140  3.147 0.2141  3.155 0.2140  3.163 0.2139  3.170 0.2139  3.177 0.2139  3.184 0.2139  3.191 0.2139  3.198 0.2139  3.205 0.2139  3.212 0.2139  3.219 0.2139  3.225 0.2139  3.231 0.2138  3.238 0.2137  3.244 0.2136  3.250 0.2136  3.256 0.2136  3.262 0.2136  3.268 0.2136  3.274 0.2136  3.280 0.2136  3.285 0.2136  3.291 0.2136  3.296 0.2136  3.302 0.2135  3.307 0.2135  3.313 0.2134  3.318 0.2133  3.323 0.2132  3.328 0.2132  3.333 0.2132  3.338 0.2132  3.343 0.2132  3.348 0.2132  3.353 0.2132  3.358 0.2132  3.362 0.2131  3.367 0.2131  3.372 0.2130  3.376 0.2129  3.381 0.2130  3.385 0.2129  3.390 0.2129  3.394 0.2129  3.399 0.2130  3.403 0.2130  3.407 0.2130  3.411 0.2130  3.416 0.2129  3.420 0.2128  3.424 0.2127  3.428 0.2127  3.432 0.2126  3.436 0.2125  3.440 0.2125  3.444 0.2125  3.448 0.2125  3.452 0.2124  3.455 0.2124  3.459 0.2124  3.463 0.2124  3.467 0.2124  3.470 0.2124  3.474 0.2124  
/

\plot 1.462 0.2137  1.732 0.2084  1.898 0.2072  2.017 0.2059  2.111 0.2047  2.188 0.2044  2.253 0.2039  2.310 0.2033  2.360 0.2029  2.405 0.2026  2.446 0.2022  2.483 0.2020  2.517 0.2017  2.549 0.2013  2.579 0.2011  2.606 0.2008  2.632 0.2004  2.657 0.2003  2.680 0.2001  2.702 0.1999  2.723 0.1997  2.744 0.1995  2.763 0.1993  2.781 0.1991  2.799 0.1989  2.816 0.1986  2.832 0.1985  2.848 0.1984  2.863 0.1983  2.877 0.1981  2.892 0.1979  2.905 0.1978  2.919 0.1977  2.931 0.1976  2.944 0.1974  2.956 0.1973  2.968 0.1972  2.980 0.1970  2.991 0.1969  3.002 0.1968  3.012 0.1967  3.023 0.1966  3.033 0.1966  3.043 0.1965  3.053 0.1965  3.062 0.1964  3.072 0.1963  3.081 0.1962  3.090 0.1961  3.098 0.1960  3.107 0.1960  3.115 0.1960  3.124 0.1960  3.132 0.1960  3.140 0.1960  3.147 0.1959  3.155 0.1958  3.163 0.1957  3.170 0.1956  3.177 0.1956  3.184 0.1956  3.191 0.1956  3.198 0.1955  3.205 0.1954  3.212 0.1954  3.219 0.1953  3.225 0.1953  3.231 0.1952  3.238 0.1950  3.244 0.1949  3.250 0.1949  3.256 0.1949  3.262 0.1948  3.268 0.1947  3.274 0.1948  3.280 0.1948  3.285 0.1947  3.291 0.1947  3.296 0.1946  3.302 0.1945  3.307 0.1944  3.313 0.1943  3.318 0.1942  3.323 0.1941  3.328 0.1941  3.333 0.1940  3.338 0.1940  3.343 0.1940  3.348 0.1939  3.353 0.1939  3.358 0.1938  3.362 0.1938  3.367 0.1937  3.372 0.1936  3.376 0.1935  3.381 0.1935  3.385 0.1934  3.390 0.1934  3.394 0.1934  3.399 0.1934  3.403 0.1934  3.407 0.1934  3.411 0.1933  3.416 0.1932  3.420 0.1932  3.424 0.1930  3.428 0.1930  3.432 0.1929  3.436 0.1928  3.440 0.1927  3.444 0.1927  3.448 0.1927  3.452 0.1925  3.455 0.1925  3.459 0.1925  3.463 0.1924  3.467 0.1925  3.470 0.1924  3.474 0.1924  
/

\plot 1.462 0.2052  1.732 0.1985  1.898 0.1962  2.017 0.1943  2.111 0.1925  2.188 0.1918  2.253 0.1909  2.310 0.1900  2.360 0.1893  2.405 0.1887  2.446 0.1881  2.483 0.1877  2.517 0.1872  2.549 0.1867  2.579 0.1862  2.606 0.1858  2.632 0.1853  2.657 0.1850  2.680 0.1846  2.702 0.1843  2.723 0.1840  2.744 0.1837  2.763 0.1834  2.781 0.1831  2.799 0.1828  2.816 0.1824  2.832 0.1822  2.848 0.1820  2.863 0.1818  2.877 0.1816  2.892 0.1813  2.905 0.1811  2.919 0.1809  2.931 0.1807  2.944 0.1805  2.956 0.1803  2.968 0.1801  2.980 0.1798  2.991 0.1797  3.002 0.1795  3.012 0.1794  3.023 0.1792  3.033 0.1791  3.043 0.1790  3.053 0.1789  3.062 0.1788  3.072 0.1786  3.081 0.1785  3.090 0.1783  3.098 0.1782  3.107 0.1781  3.115 0.1781  3.124 0.1780  3.132 0.1780  3.140 0.1779  3.147 0.1778  3.155 0.1777  3.163 0.1775  3.170 0.1774  3.177 0.1773  3.184 0.1773  3.191 0.1772  3.198 0.1771  3.205 0.1770  3.212 0.1769  3.219 0.1768  3.225 0.1767  3.231 0.1765  3.238 0.1764  3.244 0.1763  3.250 0.1762  3.256 0.1761  3.262 0.1760  3.268 0.1759  3.274 0.1759  3.280 0.1759  3.285 0.1758  3.291 0.1757  3.296 0.1756  3.302 0.1755  3.307 0.1754  3.313 0.1752  3.318 0.1751  3.323 0.1750  3.328 0.1749  3.333 0.1748  3.338 0.1748  3.343 0.1747  3.348 0.1747  3.353 0.1746  3.358 0.1745  3.362 0.1744  3.367 0.1744  3.372 0.1742  3.376 0.1741  3.381 0.1740  3.385 0.1740  3.390 0.1739  3.394 0.1738  3.399 0.1738  3.403 0.1738  3.407 0.1738  3.411 0.1737  3.416 0.1736  3.420 0.1735  3.424 0.1733  3.428 0.1732  3.432 0.1731  3.436 0.1730  3.440 0.1729  3.444 0.1728  3.448 0.1728  3.452 0.1727  3.455 0.1726  3.459 0.1726  3.463 0.1725  3.467 0.1725  3.470 0.1724  3.474 0.1724  
/

\plot 1.462 0.1968  1.732 0.1885  1.898 0.1853  2.017 0.1827  2.111 0.1804  2.188 0.1792  2.253 0.1779  2.310 0.1767  2.360 0.1757  2.405 0.1749  2.446 0.1740  2.483 0.1734  2.517 0.1727  2.549 0.1720  2.579 0.1714  2.606 0.1708  2.632 0.1701  2.657 0.1697  2.680 0.1692  2.702 0.1687  2.723 0.1683  2.744 0.1679  2.763 0.1675  2.781 0.1671  2.799 0.1667  2.816 0.1662  2.832 0.1659  2.848 0.1656  2.863 0.1653  2.877 0.1650  2.892 0.1646  2.905 0.1644  2.919 0.1641  2.931 0.1639  2.944 0.1635  2.956 0.1632  2.968 0.1630  2.980 0.1627  2.991 0.1624  3.002 0.1622  3.012 0.1621  3.023 0.1618  3.033 0.1617  3.043 0.1615  3.053 0.1613  3.062 0.1611  3.072 0.1609  3.081 0.1608  3.090 0.1605  3.098 0.1604  3.107 0.1602  3.115 0.1601  3.124 0.1600  3.132 0.1600  3.140 0.1598  3.147 0.1597  3.155 0.1595  3.163 0.1593  3.170 0.1591  3.177 0.1590  3.184 0.1589  3.191 0.1588  3.198 0.1587  3.205 0.1585  3.212 0.1584  3.219 0.1583  3.225 0.1582  3.231 0.1579  3.238 0.1578  3.244 0.1576  3.250 0.1575  3.256 0.1574  3.262 0.1572  3.268 0.1571  3.274 0.1571  3.280 0.1570  3.285 0.1569  3.291 0.1568  3.296 0.1567  3.302 0.1565  3.307 0.1564  3.313 0.1562  3.318 0.1560  3.323 0.1558  3.328 0.1558  3.333 0.1557  3.338 0.1555  3.343 0.1555  3.348 0.1554  3.353 0.1553  3.358 0.1552  3.362 0.1551  3.367 0.1550  3.372 0.1548  3.376 0.1546  3.381 0.1546  3.385 0.1545  3.390 0.1544  3.394 0.1543  3.399 0.1543  3.403 0.1542  3.407 0.1542  3.411 0.1541  3.416 0.1539  3.420 0.1538  3.424 0.1536  3.428 0.1535  3.432 0.1534  3.436 0.1532  3.440 0.1531  3.444 0.1530  3.448 0.1530  3.452 0.1528  3.455 0.1528  3.459 0.1526  3.463 0.1526  3.467 0.1526  3.470 0.1525  3.474 0.1524  
/

\color{black}
\endpicture
\caption{$Q_n(x)$ for square lattice $3$-stars as a function $\log_{10}n$ 
for $n\geq n_{min}$ and for $x=0.061565 + 0.0025\, m$ where $-4\leq m \leq 4$.  
By calculating the average slope or incline of these curves using linear fits, and 
then interpolating to find that value $x$ which gives a zero average slope or incline, 
the optimal value of $x$ at this given value of $n_{min}$ (denoted by $\xi_{n_{min}}$) 
is determined.  In this plot, $n_{min}=30$ and the optimal value of $x$ is 
$\xi_{30}\approx 0.06249$}
\label{f3}
\end{figure}

\begin{figure}[t!]
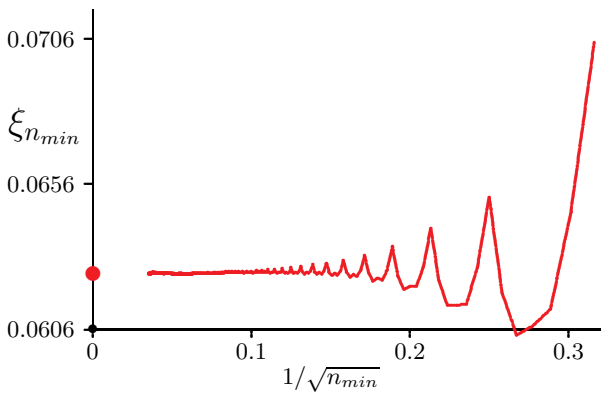

\normalcolor
\color{black}
\beginpicture
\setcoordinatesystem units <600pt,110pt>

\setplotarea x from 0 to 0.32, y from -45.5 to -44.4
\axis left shiftedto x=0 
 /
\axis bottom shiftedto y=-45.5
/

\plot 0 -45.5 -0.005 -45.5 /  \put {$0.0606$} at  -0.0325 -45.5
\plot 0 -45.0 -0.005 -45.0 /  \put {$0.0656$} at  -0.0325 -45.0
\plot 0 -44.5 -0.005 -44.5 /  \put {$0.0706$} at  -0.0325 -44.5

\plot 0 -45.5 0 -45.525 / \put {$0$} at 0 -45.575
\plot 0.1 -45.5 0.1 -45.525 / \put {$0.1$} at 0.1 -45.575
\plot 0.2 -45.5 0.2 -45.525 / \put {$0.2$} at 0.2 -45.575
\plot 0.3 -45.5 0.3 -45.525 / \put {$0.3$} at 0.3 -45.575

\put {\footnotesize$\bullet$} at 0 -45.5

\put {$1/\sqrt{n_{min}}$} at 0.15 -45.67
\put {\Large$\xi_{n_{min}}$} at -0.03 -44.8

\setplotsymbol ({\scalebox{0.3}{$\bullet$}})
\color{Red}
\plot 0.31623 -44.513157  0.30151 -45.096542  0.28868 -45.429869  0.27735 -45.488990  0.26726 -45.519497  0.25820 -45.375607  0.25000 -45.044740  0.24254 -45.287257  0.23570 -45.412270  0.22942 -45.416496  0.22361 -45.418076  0.21822 -45.330859  0.21320 -45.151598  0.20851 -45.285822  0.20412 -45.352448  0.20000 -45.353354  0.19612 -45.362234  0.19245 -45.316226  0.18898 -45.215129  0.18570 -45.295101  0.18257 -45.330550  0.17961 -45.324169  0.17678 -45.335015  0.17408 -45.308125  0.17150 -45.244650  0.16903 -45.297554  0.16667 -45.318403  0.16440 -45.310854  0.16222 -45.321251  0.16013 -45.304256  0.15811 -45.260720  0.15617 -45.299114  0.15430 -45.312807  0.15250 -45.306475  0.15076 -45.315191  0.14907 -45.302903  0.14744 -45.270982  0.14586 -45.299649  0.14434 -45.309228  0.14286 -45.303522  0.14142 -45.311245  0.14003 -45.302374  0.13868 -45.278101  0.13736 -45.300396  0.13608 -45.307415  0.13484 -45.302567  0.13363 -45.309213  0.13245 -45.302342  0.13131 -45.283635  0.13019 -45.301434  0.12910 -45.306555  0.12804 -45.302668  0.12700 -45.308457  0.12599 -45.303007  0.12500 -45.287596  0.12403 -45.302307  0.12309 -45.306243  0.12217 -45.302664  0.12127 -45.307653  0.12039 -45.303197  0.11952 -45.290437  0.11868 -45.302551  0.11785 -45.305422  0.11704 -45.302409  0.11625 -45.306866  0.11547 -45.302934  0.11471 -45.292324  0.11396 -45.302684  0.11323 -45.305023  0.11251 -45.302516  0.11180 -45.306420  0.11111 -45.303275  0.11043 -45.293988  0.10976 -45.302524  0.10911 -45.304418  0.10847 -45.302055  0.10783 -45.305499  0.10721 -45.302841  0.10660 -45.294986  0.10600 -45.302334  0.10541 -45.303791  0.10483 -45.301972  0.10426 -45.305203  0.10370 -45.303229  0.10314 -45.296214  0.10260 -45.302926  0.10206 -45.304298  0.10153 -45.302676  0.10102 -45.305609  0.10050 -45.304009  0.10000 -45.298184  0.09950 -45.304315  0.09901 -45.304247  0.09853 -45.302336  0.09806 -45.304901  0.09759 -45.303374  0.09713 -45.298138  0.09667 -45.303724  0.09623 -45.304619  0.09578 -45.303202  0.09535 -45.305594  0.09492 -45.303936  0.09449 -45.299223  0.09407 -45.303831  0.09366 -45.304466  0.09325 -45.303102  0.09285 -45.305047  0.09245 -45.303515  0.09206 -45.299228  0.09167 -45.303428  0.09129 -45.303614  0.09091 -45.302339  0.09054 -45.304240  0.09017 -45.302930  0.08980 -45.299021  0.08944 -45.303011  0.08909 -45.303567  0.08874 -45.302364  0.08839 -45.304255  0.08805 -45.302975  0.08771 -45.299878  0.08737 -45.303497  0.08704 -45.303946  0.08671 -45.302647  0.08639 -45.304429  0.08607 -45.303308  0.08575 -45.300029  0.08544 -45.303085  0.08513 -45.303672  0.08482 -45.303249  0.08452 -45.305023  0.08422 -45.304469  0.08392 -45.301643  0.08362 -45.304950  0.08333 -45.305196  0.08305 -45.304369  0.08276 -45.305874  0.08248 -45.305039  0.08220 -45.302460  0.08192 -45.305231  0.08165 -45.305740  0.08138 -45.305198  0.08111 -45.306535  0.08085 -45.305729  0.08058 -45.303515  0.08032 -45.306067  0.08006 -45.306261  0.07981 -45.305807  0.07956 -45.307362  0.07931 -45.306684  0.07906 -45.304226  0.07881 -45.306666  0.07857 -45.306958  0.07833 -45.306352  0.07809 -45.307548  0.07785 -45.307107  0.07762 -45.305186  0.07738 -45.307409  0.07715 -45.307468  0.07692 -45.306790  0.07670 -45.307909  0.07647 -45.307329  0.07625 -45.305354  0.07603 -45.307376  0.07581 -45.307588  0.07559 -45.306892  0.07538 -45.307951  0.07516 -45.307490  0.07495 -45.305625  0.07474 -45.307592  0.07454 -45.307459  0.07433 -45.306818  0.07412 -45.307701  0.07392 -45.306994  0.07372 -45.305566  0.07352 -45.307457  0.07332 -45.307222  0.07313 -45.306732  0.07293 -45.307625  0.07274 -45.306762  0.07255 -45.305138  0.07236 -45.307164  0.07217 -45.307239  0.07198 -45.306938  0.07180 -45.307485  0.07161 -45.307058  0.07143 -45.305866  0.07125 -45.307586  0.07107 -45.307465  0.07089 -45.307529  0.07071 -45.308543  0.07053 -45.308478  0.07036 -45.305843  0.07019 -45.307167  0.07001 -45.306971  0.06984 -45.306626  0.06967 -45.307448  0.06950 -45.307319  0.06934 -45.306006  0.06917 -45.307478  0.06901 -45.307345  0.06884 -45.307005  0.06868 -45.308020  0.06852 -45.307335  0.06836 -45.306308  0.06820 -45.307678  0.06804 -45.307390  0.06788 -45.306959  0.06773 -45.307741  0.06757 -45.307364  0.06742 -45.305875  0.06727 -45.307148  0.06712 -45.306872  0.06696 -45.306499  0.06682 -45.307192  0.06667 -45.306802  0.06652 -45.305819  0.06637 -45.307027  0.06623 -45.306809  0.06608 -45.306378  0.06594 -45.307519  0.06580 -45.307108  0.06565 -45.306343  0.06551 -45.307200  0.06537 -45.306982  0.06523 -45.306714  0.06509 -45.307146  0.06496 -45.306386  0.06482 -45.305671  0.06468 -45.307349  0.06455 -45.307298  0.06442 -45.307541  0.06428 -45.308410  0.06415 -45.308300  0.06402 -45.307511  0.06389 -45.308689  0.06376 -45.308394  0.06363 -45.308263  0.06350 -45.308615  0.06337 -45.308264  0.06325 -45.307483  0.06312 -45.308748  0.06299 -45.308442  0.06287 -45.308153  0.06275 -45.309081  0.06262 -45.308734  0.06250 -45.307935  0.06238 -45.309311  0.06226 -45.309546  0.06214 -45.309446  0.06202 -45.309680  0.06190 -45.309428  0.06178 -45.308805  0.06166 -45.309976  0.06155 -45.309618  0.06143 -45.309804  0.06131 -45.310560  0.06120 -45.310236  0.06108 -45.309509  0.06097 -45.310424  0.06086 -45.310005  0.06075 -45.309788  0.06063 -45.310203  0.06052 -45.309910  0.06041 -45.309463  0.06030 -45.310429  0.06019 -45.310003  0.06008 -45.309910  0.05998 -45.310442  0.05987 -45.310144  0.05976 -45.309285  0.05965 -45.310033  0.05955 -45.309715  0.05944 -45.309345  0.05934 -45.310203  0.05923 -45.310003  0.05913 -45.309071  0.05903 -45.309932  0.05893 -45.309809  0.05882 -45.309316  0.05872 -45.309646  0.05862 -45.309712  0.05852 -45.309194  0.05842 -45.309999  0.05832 -45.309478  0.05822 -45.309571  0.05812 -45.310214  0.05803 -45.310114  0.05793 -45.309611  0.05783 -45.310711  0.05774 -45.310825  0.05764 -45.311006  0.05754 -45.310073  0.05745 -45.309575  0.05735 -45.309008  0.05726 -45.309848  0.05717 -45.309724  0.05707 -45.310016  0.05698 -45.310399  0.05689 -45.310034  0.05680 -45.309322  0.05670 -45.309899  0.05661 -45.309811  0.05652 -45.309122  0.05643 -45.309744  0.05634 -45.309414  0.05625 -45.308620  0.05617 -45.309241  0.05608 -45.309090  0.05599 -45.309013  0.05590 -45.309073  0.05581 -45.308804  0.05573 -45.308031  0.05564 -45.308829  0.05556 -45.308817  0.05547 -45.308789  0.05538 -45.309499  0.05530 -45.309242  0.05522 -45.308635  0.05513 -45.309261  0.05505 -45.309605  0.05496 -45.309273  0.05488 -45.309946  0.05480 -45.309369  0.05472 -45.308656  0.05464 -45.309135  0.05455 -45.308761  0.05447 -45.308016  0.05439 -45.308737  0.05431 -45.308915  0.05423 -45.308370  0.05415 -45.309613  0.05407 -45.309468  0.05399 -45.309547  0.05392 -45.310104  0.05384 -45.309933  0.05376 -45.309171  0.05368 -45.309970  0.05361 -45.309445  0.05353 -45.309333  0.05345 -45.309659  0.05338 -45.309754  0.05330 -45.308908  0.05322 -45.309329  0.05315 -45.309460  0.05307 -45.309111  0.05300 -45.309471  0.05293 -45.309497  0.05285 -45.309382  0.05278 -45.310165  0.05270 -45.309564  0.05263 -45.309442  0.05256 -45.309999  0.05249 -45.310065  0.05241 -45.309348  0.05234 -45.310347  0.05227 -45.310404  0.05220 -45.310185  0.05213 -45.310481  0.05206 -45.310139  0.05199 -45.309345  0.05192 -45.309889  0.05185 -45.309631  0.05178 -45.309420  0.05171 -45.310163  0.05164 -45.310116  0.05157 -45.309454  0.05150 -45.310033  0.05143 -45.310036  0.05137 -45.309768  0.05130 -45.309916  0.05123 -45.309385  0.05116 -45.308843  0.05110 -45.309058  0.05103 -45.309272  0.05096 -45.309084  0.05090 -45.309057  0.05083 -45.308906  0.05077 -45.308630  0.05070 -45.308866  0.05064 -45.308489  0.05057 -45.308773  0.05051 -45.309065  0.05044 -45.309013  0.05038 -45.308210  0.05032 -45.309113  0.05025 -45.309415  0.05019 -45.309440  0.05013 -45.309863  0.05006 -45.310147  0.05000  -45.309886  0.04988 -45.308592  0.04975 -45.308354  0.04963 -45.307676  0.04951 -45.308434  0.04939 -45.308356  0.04927 -45.307791  0.04915 -45.307838  0.04903 -45.307671  0.04891 -45.307111  0.04880 -45.307478  0.04868 -45.307325  0.04856 -45.307425  0.04845 -45.307989  0.04834 -45.307722  0.04822 -45.307766  0.04811 -45.308238  0.04800 -45.307823  0.04789 -45.306736  0.04778 -45.306627  0.04767 -45.307238  0.04757 -45.307609  0.04746 -45.307977  0.04735 -45.307842  0.04725 -45.307490  0.04714 -45.307866  0.04704 -45.307827  0.04693 -45.307373  0.04683 -45.307261  0.04673 -45.308013  0.04663 -45.307217  0.04652 -45.307477  0.04642 -45.307739  0.04632 -45.308085  0.04623 -45.308396  0.04613 -45.308327  0.04603 -45.307678  0.04593 -45.308225  0.04583 -45.308005  0.04574 -45.308104  0.04564 -45.308043  0.04555 -45.307660  0.04545 -45.307031  0.04536 -45.306798  0.04527 -45.307165  0.04518 -45.306386  0.04508 -45.307162  0.04499 -45.307154  0.04490 -45.307870  0.04481 -45.308465  0.04472 -45.309416  0.04463 -45.307181  0.04454 -45.307264  0.04446 -45.307148  0.04437 -45.307169  0.04428 -45.307452  0.04419 -45.307714  0.04411 -45.307468  0.04402 -45.307328  0.04394 -45.307387  0.04385 -45.307139  0.04377 -45.307021  0.04369 -45.307920  0.04360 -45.307789  0.04352 -45.307556  0.04344 -45.308367  0.04336 -45.307916  0.04327 -45.307589  0.04319 -45.307205  0.04311 -45.306346  0.04303 -45.306859  0.04295 -45.307967  0.04287 -45.307379  0.04280 -45.307259  0.04272 -45.307911  0.04264 -45.307494  0.04256 -45.307511  0.04249 -45.307232  0.04241 -45.306425  0.04233 -45.306985  0.04226 -45.306593  0.04218 -45.306534  0.04211 -45.306810  0.04203 -45.307570  0.04196 -45.307399  0.04189 -45.307257  0.04181 -45.307238  0.04174 -45.307002  0.04167 -45.306764  0.04159 -45.307427  0.04152 -45.306901  0.04145 -45.306949  0.04138 -45.306874  0.04131 -45.306116  0.04124 -45.306475  0.04117 -45.306222  0.04110 -45.306372  0.04103 -45.306790  0.04096 -45.307590  0.04089 -45.307984  0.04082 -45.308738  0.04076 -45.306136  0.04069 -45.305974  0.04062 -45.305981  0.04056 -45.306813  0.04049 -45.306735  0.04042 -45.307267  0.04036 -45.307184  0.04029 -45.306198  0.04023 -45.306346  0.04016 -45.306463  0.04010 -45.306003  0.04003 -45.306860  0.03997 -45.306947  0.03990 -45.306324  0.03984 -45.307224  0.03978 -45.307064  0.03972 -45.306114  0.03965 -45.305546  0.03959 -45.305131  0.03953 -45.305709  0.03947 -45.306601  0.03941 -45.306498  0.03934 -45.305867  0.03928 -45.307029  0.03922 -45.307081  0.03916 -45.306689  0.03910 -45.306173  0.03904 -45.305630  0.03898 -45.305729  0.03892 -45.305327  0.03887 -45.305833  0.03881 -45.305463  0.03875 -45.306432  0.03869 -45.306506  0.03863 -45.305888  0.03858 -45.305808  0.03852 -45.305518  0.03846 -45.304921  0.03840 -45.305721  0.03835 -45.305500  0.03829 -45.304914  0.03824 -45.305435  0.03818 -45.305096  0.03812 -45.304821  0.03807 -45.304596  0.03801 -45.305274  0.03796 -45.305214  0.03790 -45.306274  0.03785 -45.307431  0.03780 -45.307853  0.03774 -45.305056  0.03769 -45.305091  0.03764 -45.305173  0.03758 -45.306024  0.03753 -45.306214  0.03748 -45.306370  0.03742 -45.306626  0.03737 -45.306296  0.03732 -45.306479  0.03727 -45.306994  0.03722 -45.306693  0.03716 -45.307137  0.03711 -45.307003  0.03706 -45.307028  0.03701 -45.307516  0.03696 -45.307203  0.03691 -45.306910  0.03686 -45.305783  0.03681 -45.305379  0.03676 -45.306766  0.03671 -45.307627  0.03666 -45.307711  0.03661 -45.307738  0.03656 -45.308540  0.03651 -45.308934  0.03647 -45.309410  0.03642 -45.308586  0.03637 -45.308236  0.03632 -45.308500  0.03627 -45.307658  0.03623 -45.308172  0.03618 -45.307827  0.03613 -45.308558  0.03608 -45.308725  0.03604 -45.308529  0.03599 -45.307904  0.03594 -45.307455  0.03590 -45.307118  0.03585 -45.307676  0.03581 -45.307500  0.03576 -45.307127  0.03571 -45.307413  0.03567 -45.307110  0.03562 -45.306880  0.03558 -45.306573  0.03553 -45.307009  0.03549 -45.307501  0.03544 -45.308217  0.03540 -45.309159  0.03536 -45.309738
 /

\put {\Large{$\bullet$}} at 0 -45.31

\color{black}
\endpicture
\caption{The estimates of $\xi_{n_{min}}$ for square lattice $3$-stars as a function of
$n_{min}$.  Notice that parity effects die down quickly.  By extrapolating to 
$n_{min}=\infty$, our best estimate of $\gamma_f{-}1$ is obtained
(denoted by the bullet on the $y$-axis).}
\label{f4}
\end{figure}

In figure \ref{f3} $Q_n(x)$ is plotted against $\log_{10}n$ for range of values of $x$.
Note that if $Q_n(x)$ is a constant, then it will present as a horisontal line in this
graph, and this will give the optimal value of $x$.  Introduce a minimum cut-off 
$n_{min}$ on the length of $f$-stars, and determine the optimal value 
$\xi_{n_{min}}$ for $x$ as described in the caption of figure \ref{f3}.  This estimate 
$\xi_{n_{min}}$ is a function of $n_{min}$.  Plotting it gives the graph in figure 
\ref{f4} (where parity effects quickly die down with increasing $n$).  
It only remains to extrapolate as $n_{min}\to\infty$.  This is done by using 
the model
\begin{equation}
\xi_{n_{min}} = (\gamma_f{-}1)  + \Sfrac{A}{\sqrt{n_{min}}} 
+ \Sfrac{B}{ n_{min} } .
\label{31x}
\end{equation}
where $250 \leq n_{min} \leq 400$.   In the case of $3$-stars this gives the
estimate $\gamma_3{-}1 \approx 0.06282$.

An error bar is determined by resampling the $\xi_{n_{min}}$.  
Each point in figure \ref{f4} is dropped with probability $0.5$ giving a smaller
set of estimates.  Extrapolating this to determine $\gamma_f{-}1$ using the
more general model 
\begin{equation}
\xi_{n_{min}} = (\gamma_f{-}1)  + \Sfrac{A}{n_{min}^{1/2}} 
+ \Sfrac{B}{ n_{min}^1 } + \Sfrac{C}{ n_{min}^{3/2} } + \Sfrac{D}{ n_{min}^2 } 
\label{32x}
\end{equation}
gives a degraded estimate of $\gamma_f{-}1$ which is dependent on the resampling
of the $\xi_{n_{min}}$.  Repeating this a large number of times gives
a distribution of estimates of $\gamma_f{-}1$ which is dependent on noise
and systematic errors in our data sets.  The variance of this distribution is 
larger than the (unknown) variance in our best estimate (since each estimate in 
the distribution is obtained by discarding data).  By taking the square root to obtain 
a standard deviation, and then doubling the standard deviation, an estimated 
error bar is obtained.  The data in figure \ref{f4} gives the estimate $\gamma_3{-}1 
= 0.06282(33)$. Repeating this analysis for the other $f$-stars gives the estimated
exponents in table \ref{t4}.

\begin{table}[h]
\caption{Estimates of $\gamma_f{-}1$ in the square lattice}
\setlength{\extrarowheight}{1.7pt}
\setlength{\tabcolsep}{8pt}
\begin{tabular}{clcc}  
\hline
 $\gamma_f{-}1$ & Exact value & This work  \cr
\hline
 $\gamma_3{-}1$ & $-0.0625 $ & $-0.06282(33)$  \cr
 $\gamma_4{-}1$ & $-0.5 $ & $-0.4989(23)\1$   \cr
 $\gamma_5{-}1$ & $-1.34375 $ & $-1.3422(13)\1$ \cr
 $\gamma_6{-}1$ & $-2.46875 $ & $-2.4668(19)\1$  \cr
 \hline
\end{tabular}
\label{t4} 
\end{table}

Equations \Ref{5} and \Ref{12} can now be used to determine the
vertex exponents $\sigma_f$ in table \ref{t1}.  We can also improve
on these estimates by using the exact value of $\sigma_1$ instead
of the estimate in equation \Ref{12}.  This gives the estimates in table \ref{t5}.

\begin{table}[h!]
\caption{Vertex exponents in 2 dimensions}
\setlength{\extrarowheight}{1.7pt}
\setlength{\tabcolsep}{8pt}
\begin{tabular} {llcr}           
\hline   
$f$& \1\1Exact & \cite{WGLW86}  & This work\scalebox{2.3}{$ $} \cr 
\hline
$\sigma_3$ & $-0.453125$     &  $-0.45(2)$& ${-0.45281(33)}$  \cr
$\sigma_4$ & $-1.1875$     & $-1.17(4)$ & ${-1.1864(23)\1}$  \cr
$\sigma_5$ & $-2.203125$   & $-2.14(4)$ & ${-2.2016(13)\1}$  \cr
$\sigma_6$ & $-3.5$ & $-3.36(5)$ & ${-3.4981(19)\1}$  \cr
\hline
\end{tabular}
\label{t5} 
\end{table}

\subsection{Estimating $\gamma_{\C{G}}{-}1$ for uniform trees}

In this section the entropic exponents $\gamma_{\C{G}}$ for 
uniform lattice trees with connectivities shown in figure \ref{f1} are estimated.
Self-avoidance in these models induces a repulsive force between nodes 
of degrees larger or equal to $3$ in uniform trees, and this stretches the
branches (self-avoiding walks) joining them.  This effect may be
more difficult to simulate with GARM, and motivated the use
of the Wang-Landau algorithm \cite{WL01} instead.  In this study we 
used a parallel implementation of this algorithm.  

We grew branched structures by first growing a central uniform star,
and then growing additional branches from the endpoints of the completed
arms of the central star.  The implementation of the Wang-Landau
algorithm grows $f$-stars by first fixing a central node.  The $f$ arms
are grown by samping $f$ edges at the endpoints of the arms and 
appending them to the star.  If it is self avoiding then the state is 
updated and accepted, and the density is updated. If it is not 
self avoiding, then the updated state is rejected, the current state 
is read again, and the density is updated accordingly.

When the star is fully grown a new (secondary) node is chosen uniformly at 
random from the $f$ endpoints of the lattice star.  Once the secondary node
is chosen the remaining branches are grown from it analogously to the arms 
of the star. 

Let $b$ denote the number of total branches (including the original 
star arms), each of length $\ell$, of the comb or brush under 
consideration. The process of first growing a star and then 
growing the remaining branches is iterated so that each structure 
of uniform length $n=b\thin \ell$ is independently sampled via the 
Wang-Landau algorithm for $\ell=1,...,200$. For each $\ell$, 
on the order of $10^9$ configurations were sampled.  The implementation
was done in parallel by growing uniform trees in separate
CPU threads using omp protocols.  These threads interacted to control 
the density update in the Wang-Landau algorithm. As with the 
parallel implementation of PERM \cite{CJvR20}, this improved 
performance of the algorithm.

As in the case of the parallel GARM sampling of stars, a symmetry
factor is introduced by the algorithm.  If $t_n(\C{G})$ is the
number of uniform square lattice combs or brushes of total length 
$n$ with $b$ arms of length $\ell$ (so that $n=b\thin \ell$), then 
the algorithm returned estimates of 
$v_n(\C{G}) = (b{-}f)!\, (f{-}1)!\,t_n(\C{G})/(4^{2f-b})$.
By equation \Ref{7},
\begin{equation}
v_n(\C{G}) 
=S_{f,b}\, t_n(\C{G}) 
=S_{f,b}\, C_{\C{G}}\,n^{\gamma_{\C{G}}-1}\,\mu_2^n\,(1+o(1))
\label{}
\end{equation}
where $S_{f,b} = (b{-}f)!(f{-}1)!/(4^{2f-b})$.
Estimates of $\gamma_{\C{G}}{-}1$ can be made by analysing these data.

\subsubsection{Estimating $\gamma_{\C{G}}{-}1$ for $\C{C}_1$, $\C{B}_1$ and
$\C{B}_2$}

We use a similar approach as for stars (see equation \Ref{15}) by 
determining the value of $x$ so that
\begin{equation}
P_n(x)=\log\left(\frac{v_{n}(\C{G})}{\mu_2^n\,n^x}\right)\simeq
C_0+C_1\,n^{-1},
\label{19}
\end{equation}
where $x=\gamma_\C{G}{-}1$.  To account for corrections due to small 
networks a cutoff $\ell_{min}$ was introduced and trees with branches of lengths
$\ell < \ell_{min}$ were excluded from the analysis.  Repeated fits for a range 
of values of $x$ gave a sequence of estimates which were interpolated to find 
that optimal value of $x$ where $P_n(x) \simeq \hbox{constant}$.  These fits 
were repeated for $\ell_{min} \in \{2,...,15\}$ and then similarly extrapolated 
against $1/\sqrt{\ell_{min}}$ to determine our best estimate of $\gamma_{\C{G}}$.

In order to determine confidence intervals on our estimates, we resampled
our data.  We selected $90\%$ of the data to estimate the exponent $x$.
This gave several data sets for each value of $\ell_{min}$.  Each of these
data sets were analysed by dropping randomly one half of the
$\ell_{min}$ and then estimating $x$.  Repeating this gave a distribution of
estimates.  Doubling the standard deviation of this distribution is our confidence 
interval.  In our particular case $90\%$ of the data were selected $10$ times and 
$50\%$ of the $\ell_{min}$ were randomly discarded 100 times.
This gave a distribution of $1,\!000$ estimates of $x$ over which the variance
was computed.  The results are shown in table \ref{t6}.   The exact values were
calculated using  the relations in equation \Ref{9}.

One may instead use the results in the last column of table \ref{t6}
to determine the $\sigma_f$ exponents using equation \Ref{9}.  This
gives the estimates in table \ref{t7} where we used the exact value
of $\sigma_1$.

The results in tables \ref{t6} and \ref{t7} shows (numerically) that the
vertex exponents $\sigma_f$, as related to uniform trees via equation \Ref{9},
are consistent.  In other words, this is strong numerical evidence that 
the results in equation \Ref{8} are correct, at least when applied to
monodisperse, acyclic, branched polymers.

\begin{table}[h!]
\caption{Estimates of $\gamma_{\C{G}}{-}1$ in the square lattice}
\setlength{\extrarowheight}{1.7pt}
\setlength{\tabcolsep}{8pt}
\begin{tabular}{clcc}  
\hline
 $\gamma_{\C{G}}{-}1$ & Exact value & This work  \cr
\hline
 $\gamma_{\C{C}}{-}1$ & $-0.21875 $ & $-0.2187(22)$  \cr
 $\gamma_{\C{B}_1}{-}1$ & $-0.78125 $ & $-0.7817(40)$   \cr
 $\gamma_{\C{B}_2}{-}1$ & $-1.34375 $ & $-1.3426(82)$ \cr
 \hline
\end{tabular}
\label{t6} 
\end{table}

\begin{table}[h!]
\caption{Vertex exponents from uniform trees}
\setlength{\extrarowheight}{1.7pt}
\setlength{\tabcolsep}{5pt}
\begin{tabular} {llcr}           
\hline   
\1\1\1$\sigma_f$& \1\1Exact & Table \ref{t5}   &  from $\gamma_{\C{G}}{-}1$  \cr 
\hline
$\sigma_3$ (via $\C{C}$) & $-0.453125$     &  $-0.45281(33)$& ${-0.4531(11)}$  \cr
$\sigma_4$ (via $\C{B}_1$) & $-1.1875$     &  $-1.1864(23)  $& ${-1.1880(51)}$  \cr
$\sigma_4$ (via $\C{B}_2$) & $-1.1875$     &  $-1.1864(23)  $& ${-1.1869(41)}$  \cr
\hline
\end{tabular}
\label{t7} 
\end{table}

\onecolumngrid

\begin{table}[h]
\caption{Estimated lattice star amplitudes in the square lattice}
\setlength{\extrarowheight}{1.7pt}
\setlength{\tabcolsep}{8pt}            
\begin{tabular}{{c} @{}*{7}{l}}
\hline                 
$f$ & $\0\0U^{(f)}$ & $\0\0C_0^{(f)}$ & $\0\0C_1^{(f)}$ 
& $\0\0C_2^{(f)}$ & $\0\0C_3^{(f)}$ & $\0\0C_4^{(f)}$ & $\0\0C_5^{(f)}$ \cr 
\hline
$1$ & $\0{1.164(23)}$  \cr
$3$ & $\0{1.2617(52)}$ & ${0.21028(86)}$ & ${0.6309(26)}$ 
& ${0.6308(26)}$  \cr
$4$ & $\0{1.2256(78)}\0$ & ${0.05107(33)}$ & ${0.2043(13)}$
& ${0.3064(20)}$ & ${0.2043(13)}$  \cr
$5$ & $\0{5.252(17)}\0$ & ${0.4377(15)}$   & ${1.3131(43)}$   
& ${0.4377(15)}$   & ${1.3131(43)}$   & ${0.8754(29)}$  \cr
$6$ & $\0{25.190(82)}$    & ${0.6997(23)}$   & ${2.0992(68)}\0$
& ${2.0992(68)}\0$  & ${0.6997(23)}$   & ${2.0992(68)}\0$ & ${2.0992(68)}\0$ \cr
\hline
\end{tabular}
\label{t8}
\end{table}

\twocolumngrid

\section{Discussion}

Accurate estimates of lattice star vertex exponents $\sigma_f$ can only 
be found if the exponent  $\sigma_1$ (and thus the entropic exponent 
$\gamma$ of self-avoiding walks) is known  with sufficient accuracy. 
We estimated $\gamma$ in equation \Ref{11},  and this compares
well with the exact value $43/32 = 1.34375$.  This, together with
our numerical results for square lattice stars, show compelling evidence
that the exact (but not rigorously proven) values of the vertex exponents 
\cite{D86,N82,N87} are correct.  In addition, our data on uniform trees 
show that the relations in equation \Ref{9} are satisfied to high accuracy,
providing evidence that the scaling relation in equation \Ref{8} is
correct as well.   Overall, our results show that the exact values of
the vertex exponents and the conjectured relations for monodispersed
acyclic branched networks in equation \Ref{9} are consistent.

In addition to estimating the vertex exponents, we also analysed our data to 
estimate the amplitudes $C$ and $C^{f}_k$ in equations \Ref{3} and \Ref{4}.
The amplitude $C$ in equation \Ref{3} is the amplitude of self-avoiding walks.
In the case of $f$-stars, $C^{(f)}_k$ is estimated taking the symmetry factors
in equations \Ref{13} and \Ref{14} into account.  Defining $U^{(f)}
= k!\,(f{-}k)!\,C^{(f)}_k$ for $3\leq f\leq 4$, and $U^{(f)}
= V^{(f)}_k\, C^{(f)}_k$ for $5\leq f \leq 6$, our data show that
asymptotically $U^{(f)}$ is independent of the parity class (see, for
example, figure \ref{f4}, where parity effects decreases quickly with increasing
$n_{min}$).

In order to estimate $U^{(f)}$, we used the log-ratio models
\begin{equation}
\log \LB \frac{u_n^{(f)}}{c_n} \RB 
= \log \LB \frac{U^{(f)}}{C} \RB + B_0 \log n + \Sfrac{C_0}{n} ,
\label{20} 
\end{equation}
and 
\begin{equation}
\log \LB \frac{u_n^{(f)}}{\sqrt{c_{2n}}} \RB 
= \log \LB \frac{U^{(f)}}{2^{\sigma_1} \sqrt{C}} \RB + B_1 \log n + \Sfrac{C_1}{n} .
\label{21}  
\end{equation}
Linear fits were done for $n\geq n_{min}$ where $n_{min} \in \{ 10,20,\ldots, 200 \}$
and the results were extrapolated using 
\begin{eqnarray}
\log \LB \frac{U^{(f)}}{C} \RB \vv_{n_{min}}
&\approx \beta_0 + \frac{\beta_1}{n_{min}} + \frac{\beta_2}{n_{min}^2} , \cr
\log \LB \frac{U^{(f)}}{2^{\sigma_1} \sqrt{C}} \RB \vv_{n_{min}}
&\approx \delta_0 + \frac{\delta_1}{n_{min}} + \frac{\delta_2}{n_{min}^2} .
\label{22}  
\end{eqnarray}

Using the estimate of $\sigma_1$ in table \ref{t1}, one can solve simultaneously for
$\{U^{(f)},C\}$.  The amplitudes $C^{(f)}_k$ are then estimated from the
value of $U^{(f)}$.  The results are shown in table \ref{t8}, where
$U^{(1)} \equiv C$ is the self-avoiding walk amplitude.  Notice that the estimates
for $f=5$ and $f=6$ appear to break the trend set by amplitudes
for $f\leq 4$.  This is due to the different style central nodes in $5$-stars 
and $6$-stars, as shown in figure \ref{f2}.

The amplitudes for the lattice networks were similarly estimated
using models like those in equations \Ref{20} and \Ref{21}, and then
extrapolated similarly to equation \Ref{22}.  Taking into account
the symmetry factors, the results in table \ref{t9} were obtained.

\begin{table}[h]
\caption{Estimates of $C(\mathcal{G})$ and $C$ in the square lattice.}
\setlength{\extrarowheight}{1.7pt}
\setlength{\tabcolsep}{10pt}       
\begin{tabular}{lcc}
\hline                  
Uniform tree & $C_{\mathcal{G}}$ &  $C$  \cr 
\hline
$\mathcal{G}=\mathcal{C}$ & $0.404(66) $  & $1.186(92) $   \cr
$\mathcal{G}=\mathcal{B}_1$ & $0.164(69) $ & $1.187(98) $  \cr
$\mathcal{G}=\mathcal{B}_2$ & $0.071(28) $ & $1.191(39) $  \cr
\hline
\end{tabular}
\label{t9}
\end{table}

\section*{Acknowledgements}
 EJJvR acknowledges financial support 
from NSERC (Canada) in the form of Discovery Grant RGPIN-2019-06303.  
SC acknowledges the support of NSERC (Canada) in the form of 
an Alexander Graham Bell Canada Graduate Scholarship 
(Application No. CGSD3-535625-2019).

\bibliographystyle{plain}
\bibliography{stars2d}

\end{document}